\documentclass[
    ,final            
  ]
  {aipproc}

\layoutstyle{6x9}

\newcommand{\NP}[1]{{ Nucl.\ Phys.\ } {\bf  #1}}
\newcommand{\ZP}[1]{{ Z.\ Phys.\ } {\bf  #1}}

\newcommand{\PL}[1]{{ Phys.\ Lett.\ } {\bf  #1}}

\newcommand{\PR}[1]{{ Phys.\ Rev.\ } {\bf  #1}}
\newcommand{\PRL}[1]{{ Phys.\ Rev.\ Lett.\ } {\bf  #1}}

\newcommand{\lsim}{\raise.3ex\hbox{$<$\kern-.75em\lower1ex\hbox{$\sim$}}}
\newcommand{\ima}{{\mbox{Im}\,}}
\newcommand{\rea}{{\mbox{Re}\,}}
\newcommand{\be}{\begin{equation}}
\newcommand{\ee}{\end{equation}}

\begin{document}

\title{Recent progress
on the light meson resonance
description from unitarized Chiral Perturbation Theory
and large $N_c$
\footnote{Invited talk to the workshop 
``Scalar Mesons: an Interesting Puzzle for QCD 
May 16 - 18,  2003, SUNY Institute of Technology, Utica, New York  }
}
\author{J.R.Pel\'aez}{address={Dipartimento di Fisica. 
Universita' degli Studi and INFN,
 Firenze,
  Italy},
  altaddress={Departamento de F\'{\i}sica Te\'orica II, 
  Universidad Complutense, 28040 Madrid, Spain}
}

\begin{abstract}
We present a brief account of the developments
in the description of light meson resonances
using unitarized extensions of the Chiral Perturbation
Theory series, both in energy and temperature. In particular,
we describe how
these methods have been recently shown to describe
simultaneously the low energy and resonance regions of
meson-meson scattering. This approach could be of
relevance to understand the light scalar mesons
since it provides a formalism that respects chiral symmetry
and unitarity and is able to generate resonant states
without any a priori theoretical bias  toward their
existence, classification or spectroscopic nature.
We will also review how this approach
is also able to describe the thermal evolution of
the $\rho$ and $\sigma$ mesons. In addition we review their 
extensions to higher orders, the most recent determination
of the resonance pole properties, as well as their behavior 
in the large $N_c$ limit, which could be of relevance to 
understand  their
spectroscopic nature.
\end{abstract}

\maketitle
\section{I. Introduction}
Although QCD has been established as the fundamental theory
of strong interactions and its predictions have been 
thoroughly tested to 
great accuracy
in the perturbative regime (above 1-2 GeV), 
our understanding of the low energy regime is still rather controversial,
specially concerning scalar mesons, which is the 
topic of this conference.
At high energies, QCD is a perturbative theory because a description
in terms of quarks and gluons is possible, however,
at low energies, the usual perturbative expansion has 
to be abandoned
in favor of somewhat less systematic approaches in terms of 
mesons. An exception
is the formalism of Chiral Perturbation Theory (ChPT) \cite{chpt1,chpt2}, 
built as the most general derivative expansion of a Lagrangian
containing {\it only} pions, kaons and the eta. These particles 
are the Goldstone bosons associated to the spontaneous
chiral symmetry breaking of massless QCD.
In practice, ChPT becomes an expansion in
powers of energy, momenta or temperature,
over the scale of the spontaneous breaking, i.e.,$4\pi f_0\simeq 1.2\,$GeV.
For the zero temperature processes
we are going to review, and due to Lorentz invariance,
only even powers of energy and momenta occur in the expansion,
which are generically denoted as $O(p^2), O(p^4)$, etc...
For thermal expansions, there is a breaking of Lorentz
invariance due to the thermal bath and the expansion
also has terms of $O(T^2), O(T^4)$ and the powers of momenta
should be separated in space and time components.
Of course, quarks are not massless, but since their mass
is small compared with typical hadronic scales, they
are introduced as perturbations 
 in the same power counting, and give rise to the 
masses of the $\pi, K$ and $\eta$ mesons, counted as $M=O(p^2)$. 
The main advantage of ChPT 
is that it provides a Lagrangian that
allows for true  Quantum Field Theory calculations, 
and a chiral power counting to organize {\it systematically}
the size of the corrections at low energies.
In particular it is possible to calculate
meson loops, whose divergences are renormalized in a finite set of
chiral parameters
at each order in the expansion.
As a consequence,  
 the parameters appearing in the Lagrangian,
after renormalization, depend on an arbitrary 
 regularization scale $\mu$;  however, the physical observables
are scale independent, since the $\mu$ dependence is canceled
through the regularization of the loop integrals. In other words,
all the relevant physical scales are those given by parameters
in the Lagrangian.
As a consequence, at each order we get finite results,
and as long as we remain at low energies and only a few
orders are necessary, the theory is
predictive in the sense that once the set of parameters
up to that order is fixed from some experiments,
it {\it should} describe, to that order,
 any other 
physical process involving mesons. 
For example, in the isospin limit, the leading order Lagrangian is universal
since it only depends on $f_0$,
which corresponds to the pion decay constant in the chiral limit,
and the leading order meson masses $M^0_\pi, M^0_K$ and $M^0_\eta$.
The dependence on the QCD dynamics only comes through the one loop
SU(3) chiral parameters  $L_i$, with $i=1...10$, 
and $H_1, H_2$. In particular, meson-meson scattering
to one loop depends only on $L_i$, with $i=1...8$ \cite{chpt2}.

 Another salient feature of ChPT
is its model independence and the fact that it has been
possible to 
obtain the behavior of the chiral parameters 
in the limit of a large number of colors $N_c$
\cite{chpt2,chptlargen}. Also, 
these parameters contain information about other heavier meson states
that have not been included as degrees of freedom in ChPT \cite{Ecker:1988te}.

As a matter of fact ChPT remains valid only at low energies, 
momenta or temperature. At higher energies
the number of independent terms allowed by symmetry
  increases dramatically  at each order, but also 
resonances appear rather soon
in meson physics. These states are
associated to poles in the second Riemann sheet of the amplitudes
and such behavior cannot be accommodated by the 
series of ChPT. From the thermal point of view, these more
massive states are always present in the bath, and their
contribution becomes relevant if the temperature increases.
Finally, a polynomial series (there are also logarithmic terms, but
are irrelevant for this discussion) will always violate the unitarity constraint,
more and more severely as the energy increases.

Thus, unitarization methods have emerged
as a powerful tool to extend the ChPT description  
with the aim of exploiting as much as possible the symmetry and 
dynamical information contained in the ChPT Lagrangian
\cite{Truong:1988zp,Dobado:1996ps,Oller:1997ng,Guerrero:1998ei,GomezNicola:2001as}. 
The basic point is to realize that the 
partial wave unitarity condition determines the 
imaginary part of the inverse of the amplitude $\ima 1/t$. 
The dynamics enter the partial waves only through $\rea 1/t$ 
and  the use of ChPT  to
calculate it has yielded remarkable results:
In \cite{GomezNicola:2001as} 
 we have recently shown 
by unitarizing the one-loop ChPT amplitudes, that
it was possible to generate the
resonant behavior of light meson resonances,
respecting simultaneously the ChPT expansion
and using values of the chiral parameters compatible
with those already present in the literature.

In what follows we will review all those results
paying special attention to the light scalar mesons.
In section II we will briefly summarize the basics 
and the most recent developments of ChPT unitarization:
the complete one-loop meson-meson scattering unitarized
within ChPT \cite{GomezNicola:2001as}, the extension to 
two-loops in $\pi\pi$ scattering and to channels
with vanishing leading order. In section III we will present 
our recent determination \cite{Coimbra} of the
pole positions of the generated resonances, which are related to
their masses and widths. 
In section IV we will introduce the thermal calculation
of $\pi\pi$ to one loop, and its unitarization that allows
for a description of the thermal behavior of the $\rho$ and $\sigma$
resonances. Finally, in section V we will also present
a study of the behavior of the poles in the large $N_c$ limit.
This whole approach is of special interest
 for the meson spectroscopy community, since these resonances are
generated from the most general Lagrangian consistent with QCD,
and therefore without any bias toward its existence, which
is still subject of a strong debate.
The fact that nine of these scalar poles seem to appear together
in chiral unitary approaches,
suggests that they could form a multiplet.
Finally, 
even more controversial is
their composition in terms of quarks and gluons,
which is properly defined only in terms of QCD.
We thus hope that the well defined link of
our approach with QCD
in the large $N_c$ limit could help shedding some light
on the spectroscopic nature of light resonances.

\section{II. Unitarized Chiral Perturbation Theory
for meson meson scattering}
In order to compare with experiment
it is customary to use partial waves $t_{IJ}$ of
definite isospin $I$ and angular momentum $J$. 
For simplicity we will omit 
the $I,J$ subindices in what follows, so that the chiral expansion becomes
$t\simeq t_2+t_4+...$, with $t_2$ and $t_4$ of ${O}(p^2)$ and ${O}(p^4)$, respectively. 
The unitarity relation  for the partial waves $t_{ij}$,
where $i,j$ denote the different available states, is very simple: 
when two states, say "1" and ``2'', are accessible,
it becomes
\be
\ima T = T \, \Sigma \, T^* \quad \Rightarrow \quad \ima T^{-1}=- \Sigma
\quad  \Rightarrow \quad T=(\rea T^{-1}- i \,\Sigma)^{-1}
\label{unimatrix}
\ee
with \vspace*{-.5cm}
\be
T=\left(
\begin{array}{cc}
t_{11}&t_{12}\\
t_{12}&t_{22} \\
\end{array}
\right)
\quad ,\quad
\Sigma=\left(
\begin{array}{cc}
\sigma_1&0\\
0 & \sigma_2\\
\end{array}
\right)\,,
\ee
where $\sigma_i=2 q_i/\sqrt{s}$ and $q_i$ is the C.M. momentum of
the state $i$. The generalization to more than two accessible
states is straightforward in this matrix notation.
Let us remark that, since $\ima T^{-1}$ is fixed by unitarity
{\it we only need to know the real part of the Inverse Amplitude}.
Note that Eq.(\ref{unimatrix}) is non-linear and cannot be satisfied 
exactly with a perturbative expansion like that of ChPT, although
it holds perturbatively, i.e,
\vspace*{-.2cm}
\begin{eqnarray}
\ima T_2 = 0, \quad \quad \ima T_4 = T_2 \, \Sigma
\, T_2^*\,+ {O}(p^6) . \label{pertuni}
\end{eqnarray}
The use of 
the ChPT expansion 
$\rea T^{-1}\simeq  T_2^{-1}(1-(\rea T_4) T_2^{-1}+...)$ 
in eq.(\ref{unimatrix}),
guarantees that we reobtain the ChPT low energy expansion
and that we are taking into account all the information
included in the chiral Lagrangians (both about $N_c$ and 
about heavier resonances).
In practice, all the powers of $1/f_0$ in the amplitudes
are rewritten 
in terms of physical constants $f_\pi$ or $f_K$ or $f_\eta$.
At leading order this difference is irrelevant, but at 
one loop, we have three possible choices for each power of 
$f_0$ in the amplitudes, all equivalent up to ${O}(p^6)$. 
It is however possible to substitute the $f_0$'s by
their expressions in terms of $f_\pi$ or $f_K$ or $f_\eta$
in such a way that they cancel the ${ O}(p^6)$ and higher order
contributions in eq.(\ref{pertuni}), so that
\vspace*{-.2cm}
\begin{eqnarray}
\ima T_2 = 0, \quad \quad \ima T_4 = T_2 \, \Sigma
\, T_2^*. \label{exactpertuni}
\end{eqnarray}
We will call these conditions ``exact perturbative unitarity''.
From  eq.(\ref{exactpertuni}),
we find
\begin{equation}
 T\simeq T_2 (T_2-T_4)^{-1} T_2,
\label{IAM}
\end{equation}
which is the coupled channel IAM
that has been used 
to unitarize simultaneously all the one-loop ChPT meson-meson
scattering amplitudes \cite{GomezNicola:2001as}. It has the advantage
that all the pieces are analytic and it is thus straightforward
to obtain analytic continuations in the complex $s$ plane 
to look for poles associated to resonances. 
Although the justification of the IAM we have presented is valid
only for physical values of $s$, where the unitarity condition holds,
the analytic continuation 
to the complex $s$ plane has also been justified in terms 
of dispersion relations in the elastic case 
\cite{Truong:1988zp,Dobado:1996ps}. One of the main advantages of the IAM
is that it is extremely  simple to implement, only involving 
algebraic manipulations on the ChPT series. Alternative methods
have been proposed and applied successfully to the full ChPT series, 
for instance for $\pi\pi$ scattering to one loop \cite{Nieves:1999bx}.
However, in this brief review
we concentrate on the IAM 
only due to its simplicity and remarkable success.

The IAM was applied first for 
partial waves in the elastic region, where a single channel
is enough to describe the data. 
This approach was able to generate
the $\rho$ and $\sigma$ poles
in $\pi\pi$ scattering and that of  $K^*$ in $\pi K\rightarrow\pi K$
\cite{Dobado:1996ps}. 
Only later it has been noticed that the $\kappa$ pole can also be
obtained in the elastic single channel formalism.
Concerning coupled channels, since
not all the meson-meson amplitudes were known to one-loop,
only the leading order and the dominant s-channel loops
were considered in \cite{Oller:1997ng}, thus
neglecting crossed and tadpole loop
diagrams. Despite these approximations, it was achieved
a remarkable description of meson-meson
scattering up to 1.2 GeV, generating  the poles associated to the 
$\rho$, $K^*$, $f_0$, $a_0$, $\sigma$ and $\kappa$.
The price to pay was, first,  that only 
the leading order of the expansion
was recovered at low energies. Second, apart from the fact
that loop divergences  were regularized with a cutoff, 
thus introducing an spurious parameter,
they 
were not completely renormalized, since
there were diagrams missing. Therefore, 
it was not possible to compare the $L_i$ parameters,
which are supposed to encode the underlying QCD dynamics, 
with those already present in the literature.

 As already explained
the whole approach is rather interesting to study the existence 
and properties of 
light scalar resonances and 
it is then very relevant to check that
these poles and their features
are not just artifacts of the approximations,
estimate the uncertainties in their parameters, and
check their compatibility with other experimental information
regarding ChPT. All in all, it is also 
hard to study within this approximation the large $N_c$ behavior 
that should be inherited by ChPT from QCD.

The above reasons triggered the interest
in calculating and unitarizing the remaining meson-meson amplitudes
within one-loop ChPT. In  \cite{Guerrero:1998ei}
the
$K\bar{K}\rightarrow K\bar{K}$ calculation
was completed, and  together with previous
works \cite{Kpi}, they
allowed for the unitarization of the $\pi\pi$, $K\bar{K}$
coupled system. There was a good agreement
of the IAM description with the existing $L_i$, reproducing
again the resonances in that system.
Much more recently, we have completed  
the one-loop meson-meson scattering calculation \cite{GomezNicola:2001as},
including three new amplitudes:
$K\eta\rightarrow K\eta$, $\eta\eta\rightarrow\eta\eta$ and
$K\pi\rightarrow K\eta$, but recalculating
the other five amplitudes unifying the  notation, ensuring
`` exact perturbative
unitarity'', Eq.(\ref{exactpertuni}), 
and also correcting some errors in the literature.
Next, we have unitarized these amplitudes with the coupled channel IAM 
thus allowing for a direct comparison with the
standard low-energy chiral parameters, in very good
agreement with previous determinations. 
In that work 
we presented the full calculation of all the one-loop
amplitudes in dimensional regularization, 
and a 
simultaneous description of
  the low energy and the resonance regions. In addition we
estimated the uncertainties from the data, which are rather large 
due to the intrinsic difficulties in the meson-meson experiments.

Obviously, the first check was to use the  
standard ChPT parameters, that we have given in Table 1
to see if the resonant features were still there, at least qualitatively,
and they are. Thus, they are not just an artifact of the approximations 
and of the values chosen for the parameters.
As already commented, this comparison can only be performed now since
we have all the amplitudes
renormalized in the standard $\overline{MS}-1$ scheme.

\begin{table}[hbpt]
\begin{tabular}{|c||c|c||c|c|c|}
\hline
  \tablehead{1}{r}{b}{Parameter} &
  \tablehead{1}{c}{b}{
$K_{l4}$ decays} &
  \tablehead{1}{c}{b}{\hspace*{0.6cm}ChPT\hspace*{0.6cm} } &
  \tablehead{1}{c}{b}{\hspace*{0.6cm}IAM I\hspace*{0.6cm}}&  
  \tablehead{1}{c}{b}{\hspace*{0.6cm}IAM II\hspace*{0.6cm}} &
  \tablehead{1}{c}{b}{\hspace*{0.6cm}IAM III\hspace*{0.6cm}} 
\\
\hline
$L_1^r(M_\rho)$
& $0.46$
& $0.4\pm0.3$
& $0.56\pm0.10$ 
& $0.59\pm0.08$
& $0.60\pm0.09$
\\
$L_2^r(M_\rho)$
& $1.49$
& $1.35\pm0.3$ 
& $1.21\pm0.10$ 
& $1.18\pm0.10$
& $1.22\pm0.08$\\
$L_3 $  &
 $-3.18$ &
 $-3.5\pm1.1$&
$-2.79\pm0.14$ 
&$-2.93\pm0.10$
& $-3.02\pm0.06$
\\
$L_4^r(M_\rho)$
& 0 (fixed)
& $-0.3\pm0.5$& $-0.36\pm0.17$ 
& $0.2\pm0.004$
& 0 (fixed)\\
$L_5^r(M_\rho)$
& $1.46$
& $1.4\pm0.5$& $1.4\pm0.5$ 
& $1.8\pm0.08$
& $1.9\pm0.03$
\\
$L_6^r(M_\rho)$
& 0 (fixed)
& $-0.2\pm0.3$& $0.07\pm0.08$ 
&$0\pm0.5$
&$-0.07\pm0.20$\\
$L_7 $  & $-0.49$ & 
$-0.4\pm0.2$&
$-0.44\pm0.15$ &
$-0.12\pm0.16$&
$-0.25\pm0.18$
\\
$L_8^r(M_\rho)$
& $1.00$
& $0.9\pm0.3$& $0.78\pm0.18$ 
&$0.78\pm0.7$
&$0.84\pm0.23$\\
\hline
\end{tabular}
\caption{Different sets of chiral parameters ($\times10^{3}$).
The first column comes from recent analysis of $K_{l4}$ decays
\cite{BijnensKl4} ($L_4$ and $L_6$ are set to
zero). In the ChPT column $L_1,L_2,L_3$ come from
\cite{BijnensGasser} and  the rest from  \cite{chpt2}. The 
three last 
ones correspond to the values from the IAM including the
uncertainty due to different systematic error used on different
fits. Sets II and III are obtained using amplitudes 
expressed in terms of $f_\pi$, $f_K$ and $f_\eta$, whereas
the amplitudes in set I are expressed in terms of $f_\pi$ only.} \label{eleschpt}
\end{table}

After checking that we made an IAM fit.  
Systematic errors
in the data are the largest contribution to the
error bands of the results 
as well as in the fit parameters in Table 1.
These uncertainties 
are calculated from a MonteCarlo Gaussian sampling \cite{GomezNicola:2001as}
(1000 points)
of the $L_i$ sets within their error bars, assuming they are
uncorrelated. Note that in Table 1 we list three sets of
parameters for the IAM fit, fairly compatible among them and with 
those of standard ChPT. As explained before
they correspond to different choices when reexpressing
the $f_0$ parameter of the Lagrangian in terms of
physical decay constants.
The IAM I fit was obtained using just $f_\pi$,
which is simpler but unnatural 
when dealing with kaons or etas.  
The plots and uncertainties of this fit are not shown here
because they can be found in \cite{GomezNicola:2001as}.
There, it could be observed that the $f_0(980)$  region
was not very well described yielding a
too small width for the resonance.

That is the reason why Fig.1 shows the results of
a second fit (IAM II) using
amplitudes written in terms of 
$f_K$ and $f_\eta$ when dealing with processes
involving kaons or etas \cite{Coimbra} .
Let us remark that the data in the $f_0(980)$ region
is well within the uncertainties.
In particular, we have rewritten
our $O(p^2)$ amplitudes replacing one factor of $1/f_\pi$ by
$1/f_K$ for each two kaons present
between the initial or final state, or by $1/f_\eta$ 
for each two etas appearing between the initial and final states.
In the special case $K\eta\rightarrow K\pi$ we have changed 
$1/f_\pi^2$ by $1/(f_Kf_\eta)$.
The difference between the two ways of writing
the leading order amplitudes is $O(p^4)$, and is therefore included
in the next to leading order contribution using the relations between
the decay constants and $f_0$ provided in \cite{chpt2,GomezNicola:2001as}. 
The $1/f_0$ factors in each loop function  at $O(p^4)$
(generically, the $J(s)$
given in the appendix of \cite{GomezNicola:2001as})
are changed to satisfy ``exact perturbative unitarity'',
eqs.(\ref{exactpertuni}). From the point
of view of the ChPT counting, the amplitudes are the same 
up to $O(p^4)$, but numerically they are slightly different.
From Table 1, we see that 
the only sizable change is in the parameters related to the 
decay constants, i.e., $L_4$ and $L_5$. 
For illustration we give in Table 1 a third fit, IAM III, 
obtained as IAM II but fixing $L_4=0$ as in 
the most recent $K_{l4}$  $O(p^4)$ 
determinations. This is the value 
estimated from the large $N_c$ limit,
and since our fits are not very sensitive to the 
variations in $L_4$ thus we avoid that it could get an unnatural value 
just from an insignificant improvement
of the $\chi^2$.
\begin{figure}[hbpt]
\includegraphics[height=0.7\textheight]{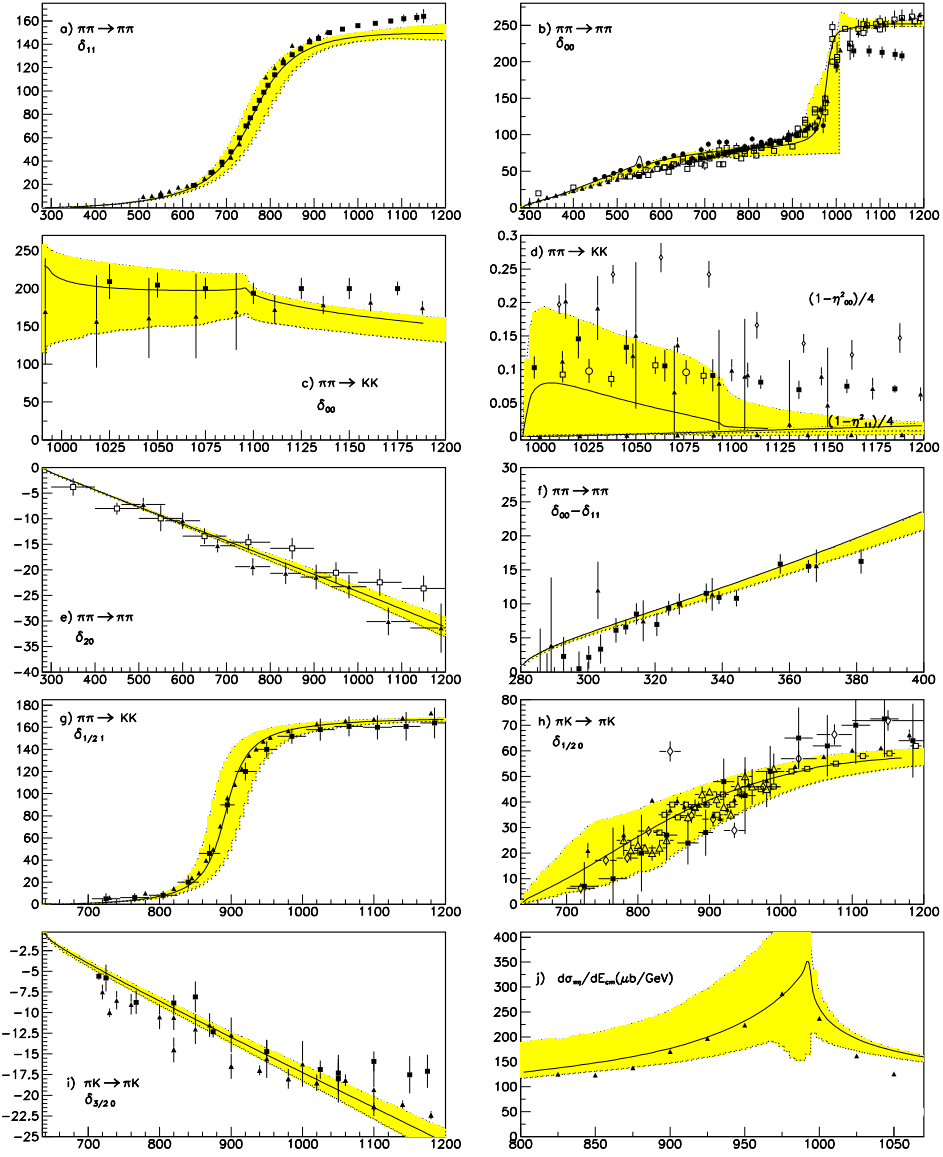}
\caption
{ IAM fit to meson-meson scattering data, set II in Table 1.
The uncertainties cover  the estimated systematic errors.
The statistical errors from the fit would be much smaller.
The data comes from \cite{pipidata}} 
\label{fig:Tps1}
\end{figure}

\begin{table}[h]
\begin{tabular}{|c|c|c|c|c|}
\hline 
Threshold&Experiment
&\hspace{0.5cm}IAM fit I\hspace*{0.5cm}&\hspace{0.5cm}ChPT ${O}(p^4)\hspace{0.5cm}$
&\hspace*{0.5cm}ChPT ${O}(p^6)$\hspace*{0.5cm}\\
parameter&&\cite{GomezNicola:2001as}&\cite{Dobado:1996ps,Kpi}
&\cite{Amoros:2000mc}\\ 
\hline \hline 
$a_{0\,0}$&0.26 $\pm$0.05&0.231$^{+0.003}_{-0.006}$&0.20&0.219$\pm$0.005\\
$b_{0\,0}$&0.25 $\pm$0.03&0.30$\pm$ 0.01&0.26&0.279$\pm$0.011\\
$a_{2\,0}$&-0.028$\pm$0.012&-0.0411$^{+0.0009}_{-0.001}$&-0.042&-0.042$\pm$0.01\\
$b_{2\,0}$&-0.082$\pm$0.008&-0.074$\pm$0.001&-0.070&-0.0756$\pm$0.0021\\
$a_{1\,1}$&0.038$\pm$0.002&0.0377$\pm$0.0007&0.037&0.0378$\pm$0.0021\\ 
$a_{1/2\,0}$&0.13...0.24&0.11$^{+0.06}_{-0.09}$&0.17&\\
$a_{3/2\,0}$&-0.13...-0.05&-0.049$^{+0.002}_{-0.003}$&-0.5&\\
$a_{1/2\,1}$&0.017...0.018&0.016$\pm$0.002&0.014&\\
$a_{1\,0}$&&0.15$^{+0.07}_{-0.11}$&0.0072&\\ 
\hline
\end{tabular}
\caption{ Scattering lengths $a_{I\,J}$ and slope parameters
$b_{I\,J}$ for different meson-meson scattering channels. For
experimental references see \cite{GomezNicola:2001as}. Let us
remark that our one-loop IAM results at threshold are very similar
to those of two-loop ChPT.}\label{elesfit}
\end{table}

To conclude, we show in  Table 2 the 
values of the scattering 
lengths and slopes, which confirms that we have
a simultaneous description of the low energy and resonant
regions
of meson-meson scattering. As we have seen in Table 1, all this is achieved 
with chiral parameters  compatible with those from standard ChPT.
Hence, since the expressions
are fully renormalized, we are not including any dependence on any  
spurious parameter.

Another result of relevance in the context of
unitarized ChPT has been the consideration 
of higher order effects. There is indeed a simple way to 
extend the IAM to higher orders, already proposed in \cite{Dobado:1996ps},
and first applied to two loop $\pi\pi$ scattering and the
pion form factor \cite{Hannah:1997ux},
also obtaining remarkable results.
This study has been very recently extended in \cite{Nieves:2001de} 
with a careful analysis of the uncertainties. This amplitude depends
on the  $O(p^4)$ and $O(p^6)$ parameters through six combinations
only, called $b_i$.
Despite the poor knowledge about these two-loop parameters,
listed in Table 3,
in  \cite{Nieves:2001de} a good description of the data is found, 
including the $\sigma$ and $\rho$ regions,
with parameters, again given in Table 3, compatible within errors with those 
already in the literature. The error analysis carried out in 
\cite{Nieves:2001de} is also of relevance because 
unitarization methods in
general, and the IAM in particular, are mainly criticized for their
violation of  crossing symmetry. However, in that work it was shown that 
the IAM
crossing violations, ``quantified in terms 
of Roskies sum rules'' , and {\it taking into account
the present
experimental uncertainties} are `` not very large in percentage terms''.

We finish this section about the unitarized description
of meson-meson scattering reviewing the recent generalization
of the IAM \cite{Dobado:2001rv}
to channels where the leading order, $T_2$, vanishes.
For instance, that is the case for all channels with $J>1$,
so that eq.(\ref{IAM}) cannot be applied.
Note also that eq.(\ref{pertuni}) implies that the 
perturbative imaginary part vanishes up to $O(p^8)$. 
However, despite these difficulties, and using again a dispersive approach,
it has been possible to generalize the IAM 
and unitarize the $(2,0)$ channel, using:
\begin{equation}
t^{IAM}=\frac{t_4}{1-t_6/t_4-t_8/t_4+t_6^2/t_4^4}.
\label{j2}
\end{equation}
Up to the moment not even  the $O(p^8)$ $SU(2)$ pion ChPT
Lagrangian is known. Hence, we have used the two-loop $SU(2)$ calculation
and we have estimated the $O(p^8)$ contribution to $\pi\pi$ scattering 
in the chiral limit, since, as it is the only term surviving
in the $m_\pi\rightarrow0$ limit, we expect it to dominate
at the energies $\sqrt{s}>>m_\pi$
 where the first $f_2(1250)$ resonance appears.
In this way we are only introducing an additional parameter, 
whose size,  on dimensional grounds,
should be $c\simeq(1/4\pi f_0)^8\simeq 3\times10^{-25}\hbox{MeV}^{-8}$

\begin{table}[hbpt]
    \begin{tabular}{|c|c|c|c|c|c|c|}
\hline
&$10^2b_1$ & $10^2b_2$ & $10^3b_3$ & $10^3b_4$ & $10^4b_5$ & $10^4b_6$ \\ \hline
$O(p^6)$ ChPT \cite{bijnens}& -9.2 ... -8.6 & 8.0 ... 8.9 & -4.3 ... -2.6 & 4.8 ... 7.1 & -0.4 ... 2.3 & 0.7 ... 1.5 \\ 
$O(p^6)$ IAM \cite{Nieves:2001de} &$-7.7\pm1.3$ & $7.3\pm0.7$ &$-1.8\pm1.6$ & $4.8\pm0.1$ & $1.3\pm0.2$ &$0.2\pm0.2$ \\
set I &-3 &4 & 3.8 &7 & 8.7 & 1.6 \\
set II &  -6.6 & 6.4 & -3.6 & 6.7 & 4.0 & 1.5 \\\hline
    \end{tabular}
    \caption{ Estimates of the $O(p^6)$ parameters
are given in row two. In the third row we
 give values that,
with the IAM up to $O(p^6)$, fit very well 
$\pi\pi$ scattering in
the $(I,J)= (0,0), (1,1)$ and $(2,0)$ channels.
Set I with eq.(8)
describes remarkably well the $I=0,J=2$ data, but only agrees 
in the order of magnitude with previous values. Set II 
is closer to \cite{Nieves:2001de}, but yields a narrower resonance
(see Fig.2.a), due to other coupled states not present
in our approach. }
\end{table}

Thus, in Fig.2.a (dashed line) we compare the $(I,J)=(0,2)$
phase shift data with the results of applying eq.(\ref{j2}) to the
ChPT amplitude just described, using the set I of parameters given in Table 3.
By fitting, it is possible to force a 
remarkable description of the experiment, including the $f_2(1250)$
resonance. However that is not realistic since the $f_2(1250)$
has an $85\%$ decay to  $\pi\pi$ and should not be 
described with  $\pi\pi$ scattering only. For that reason the set I,
although with the correct order of magnitude,
does not compare well with the values given in the literature, 
listed in column two
and three of Table 3. 
For illustrative purposes, we also show,
as a dotted line, the result using set II in Table 3, which is
obtained if we allow for a $ 15\% $ narrower resonance.
That is not a fit, but the parameters are much closer to
those given in the literature.  Finally, we want to remark
that for any set
of parameters that we have found yielding results like those 
in Fig.2.a, 
the values needed for the $c$ constant lie in the $c\simeq 10^{-25}$
to $10^{-24}\,$ MeV$^{-8}$ range, consistently with our expectations.

We have therefore checked that the IAM produces good results
as soon as we provide it with reasonable inputs, also in channels
with higher angular momenta.

\begin{figure}[hbpt]
\hspace*{.3cm}
\includegraphics[width=0.34\textheight]{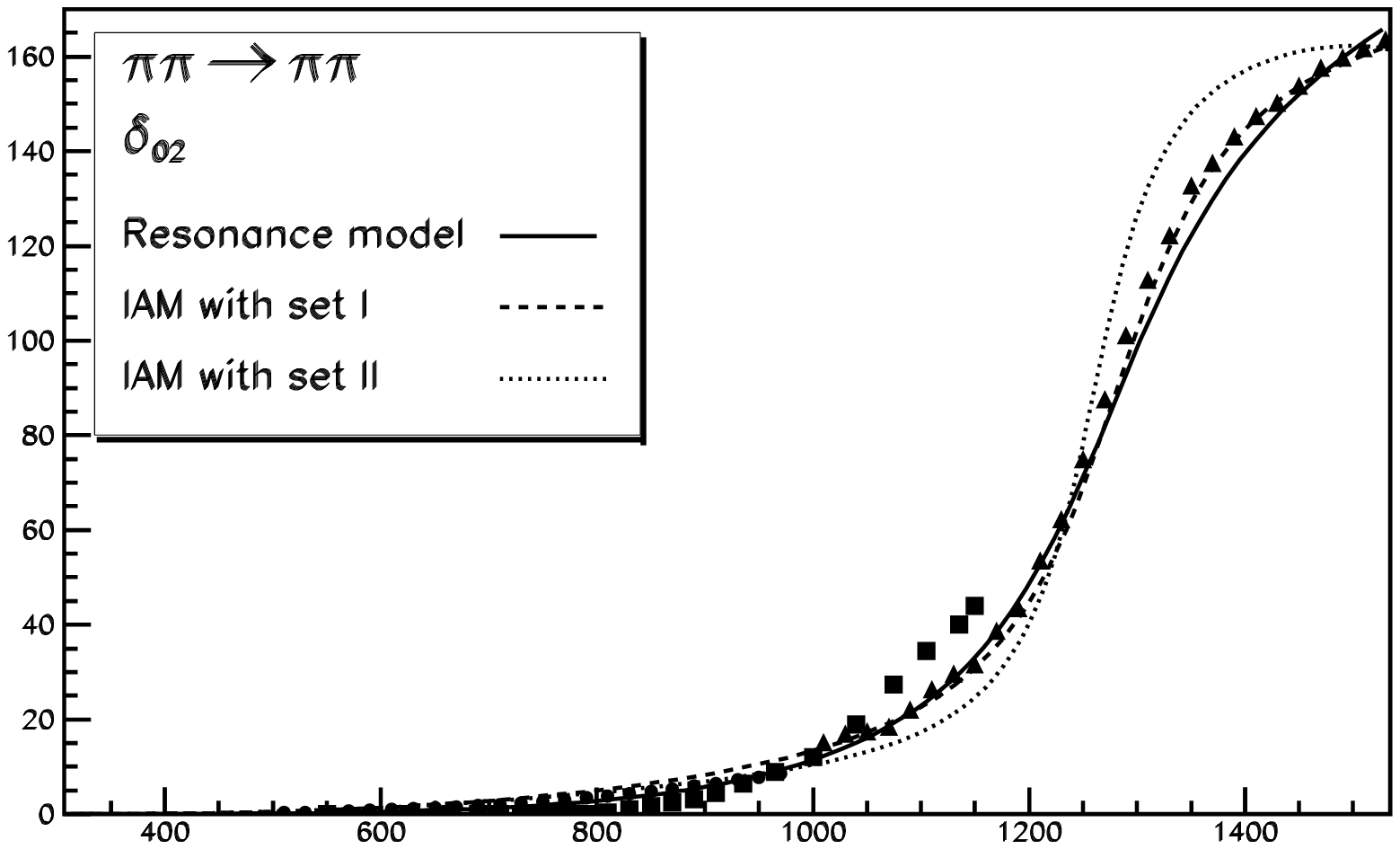}
\includegraphics[width=0.35\textheight]{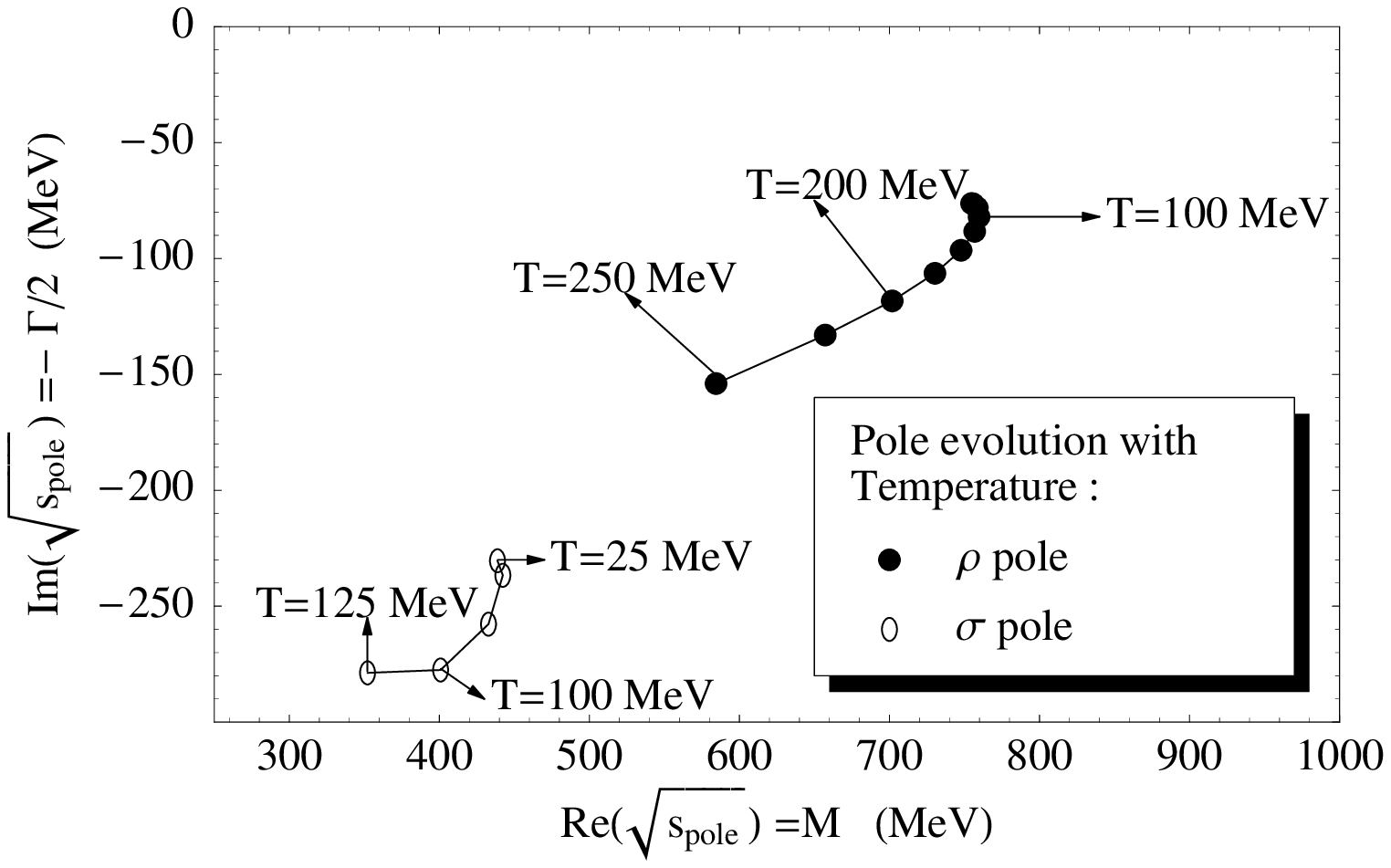}
\caption
{a) The $,I=0, J=2$ $\pi\pi$ scattering
phase shift \cite{pipidata} and the results of
our unitarized chiral resonance model (continuous line), 
of eq.(\ref{j2})
with the parameters in set I (dashed line), and with those 
in set II (dotted line) in Table 3.  b) Thermal evolution
of the $\sigma$ and $\rho$ poles.}
\label{fig:f2}
\end{figure}

\section{IV. Poles associated to resonances}

In this section we will investigate whether 
the unitarized complete meson-meson  amplitudes 
obtain the same poles as obtained with previous approaches.
This is of relevance since, for instance, both the
$\sigma$ and $\kappa$ scalar states are rather controversial
even though their poles have been found 
by several other groups using different techniques \cite{newsigma,JA}. 
The debate has become increasingly interesting when 
recent experiments seem to require such poles \cite{charm}
to describe completely different processes.

The most interesting features of the chiral unitary
approaches is that the poles thus generated
are not included in the original ChPT Lagrangian and hence
appear without any theoretical prejudice toward
their existence, classification in multiplets, or nature.
Of particular interest  for this workshop is 
the simultaneous generation of 
the scalar resonances $\sigma, \kappa, a_0(980)$ and $f_0(980)$
in a chiral context, so that it seems very
natural to interpret them as a nonet.
Nevertheless, we should distinguish two 
different resonance generation mechanisms:
already in \cite{Oller:1997ti} it was noted that to generate
the scalars just the leading order and
a cutoff was enough, whereas the vector mesons
require the chiral parameters, particularly $L_1, L_2$ and $L_3$
\cite{Oller:1997ng}. Of course, the chiral parameters are always present ,
but it seems that their values are {\it not} related 
to the hypothesized light scalar nonet.
Since the vectors are  fairly well established $q\bar{q}$ states,
this difference suggests that scalars like the $\sigma$, $\kappa$, etc
may have a different nature. With the amplitudes described in the previous
sections we expect to reach a more conclusive statement,
since they respect the chiral expansion, and, 
being completely renormalized, have no
spurious dependence on any cutoff or 
dimensional regularization scale. In addition, the fact
that we use chiral parameters compatible with previous determinations
ensures that our description is consistent with other low
energy processes.

Thus, in Table 4 we show the pole position for the resonances, 
including  uncertainties,
for the different IAM parameter sets given in Table 1.
For comparison we provide in the first
row those obtained
in the ``approximated'' IAM \cite{Oller:1997ng}, 
whereas we list in Table 5 the poles listed presently in the
PDG \cite{PDG}. These results deserve some comments:
\begin{itemize}
\item The vectors $\rho(770)$ and $K^*(892)$, are very stable
within chiral unitary approaches. Their positions are almost
the same irrespective of whether one uses the single channel 
\cite{Truong:1988zp,Dobado:1996ps}, 
the approximated coupled channel \cite{Oller:1997ng}
or the complete IAM \cite{GomezNicola:2001as}.

\item The $\sigma$ and $\kappa$ pole positions are robust
within these approaches. No matter what version
of the IAM is used. Note the small uncertainties in some of their parameters,
in very good agreement with recent experiments \cite{charm}.

\item  The $f_0(980)$  has  a sizable
 decay to two different channels and therefore it can only be studied
in a coupled channel formalism. In practice, both the 
approximated and complete IAM generate a pole
associated to this state at approximately the same mass.
However, as already remarked the unitarized ChPT amplitudes 
using just $f_\pi$ \cite{GomezNicola:2001as}, yield a 
too narrow width. The good news is that it can be well 
accommodated using $f_\pi$, $f_K$ and $f_\eta$ \cite{Coimbra}.

\item   The $a_0(980)$ also requires a coupled channel formalism,
and the data on this region is well described either by the approximated or
the complete IAM. However, the presence of a pole is 
strongly dependent on whether we write the ChPT amplitudes
only in terms of $f_\pi$, or using the three $f_\pi$, $f_K$ and $f_\eta$.
In the first case, as
already pointed out in \cite{Uehara:2002nv},  the use of the
``approximated'' IAM with just $f_\pi$, favored 
a ``cusp'' interpretation
of the $a_0(980)$ enhancement in $\pi\eta$ production. 
With the complete IAM we do not find a pole near the
$a_0(980)$ enhancement and indeed the $\pi\eta$ phase-shift
does not cross $\pi/2$ and it neither has a fast phase movement.
In contrast, when expressing $f_0$ in terms of
 $f_\pi$, $f_K$ and $f_\eta$ as described in the previous section,
we do find a pole and its associated fast phase movement through $\pi/2$,
either with the approximated or complete IAM. 
Thus, this pole is rather unstable as can 
be noticed from its large uncertainties in Table 4. We are 
somewhat more
favorable toward the second interpretation because it is also able to 
describe better the $f_0(980)$ width. But the two possibilities 
remain open.
\end{itemize}

\begin{table}[htbp]
\begin{tabular}{ccccccc}
\hline
$\sqrt{s_{pole}}$(MeV)
&$\rho$
&$K^*$
&$\sigma$
&$f_0$
&$a_0$
&$\kappa$
\\ \hline
$^{\hbox{IAM Approx}}_{\;\;\;
\hbox{(no errors)}}$
&759-i\,71
&892-i\,21
&442-i\,227
&994-i\,14
&1055-i\,21
&770-i\,250
\\\hline
IAM I
&760-i\,82
&886-i\,21
&443-i\,217
&988-i\,4
& cusp?
&750-i\,226
\\
(errors)
&$\pm$ 52$\pm$ i\,25
&$\pm$ 50$\pm$ i\,8
&$\pm$ 17$\pm$ i\,12
&$\pm$ 19$\pm$ i\,3
&
&$\pm$18$\pm$i\,11
\\ \hline
IAM II
&754-i\,74
&889-i\,24
&440-i\,212
&973-i\,11
&1117-i\,12
&753-i\,235
\\
(errors)
&$\pm$ 18$\pm$ i\,10
&$\pm$ 13$\pm$ i\,4
&$\pm$ 8$\pm$ i\,15
&$^{+39}_{-127}$ $^{+i\,189}_{-i\,11}$
&$^{+24}_{-320}$ $^{+i\,43}_{-i\,12}$
&$\pm$ 52$\pm$ i\,33\\\hline
IAM III
&748-i68
&889-i23
&440-i216
&972-i8
&1091-i52
&754-i230
\\
(errors)
&$\pm$ 31$\pm$ i\,29
&$\pm$ 22$\pm$ i\,8
&$\pm$ 7$\pm$ i\,18
&$^{+21}_{-56}$$\pm$ i\,7
&$^{+19}_{-45}$ $^{+i\,21}_{-i\,40}$
&$\pm$ 22$\pm$ i\,27\\
\hline
\end{tabular}
\caption{ Pole positions (with errors) in meson-meson scattering.
When close to the real axis the mass and width of the 
associated resonance is $\sqrt{s_{pole}}\simeq M-i \Gamma/2$.}
\end{table}

\begin{table}[htbp]
\begin{tabular}{ccccccc}
\hline
PDG2002
&$\rho(770)$
&$K^*(892)^\pm$
&$\sigma$ or $f_0(600)$
&$f_0(980)$
&$a_0(980)$
&$\kappa$
\\ \hline
Mass (MeV)
&$771\pm0.7$
&$891.66\pm0.26$
&(400-1200)-i\,(300-500)
&$980\pm10$
&$980\pm10$
&not\\
Width (MeV)
& $149\pm0.9$
& $50.8\pm0.9$
&(we list the pole)
&40-100
&50-100
&listed\\\hline
\end{tabular}
\caption{ Mass and widths or pole positions 
of the light resonances quoted in the PDG \cite{PDG}.
Recall that for narrow resonances $\sqrt{s_{pole}}\simeq M-i \Gamma/2$}
\end{table}

Let us remark that the $f_0(980)$ and $a_0(980)$ resonances
are very close to the $K\bar{K}$ threshold, which can induce a considerable
distortion in the resonance shape, whose relation to the pole position
could be far from that expected for narrow resonances. In addition
these states have a large mass and it is likely that their nature 
should be understood from a mixture with heavier states.

\section{IV. Thermal evolution of the $\sigma$ and $\rho$ mesons.}

In a recent paper we have calculated the temperature  one-loop 
corrections to the $\pi\pi$ scattering amplitude \cite{GomezNicola:2002tn}. 
These corrections are finite although they appear through 
the pion loops and consequently, do not include any additional
parameter in the expression of the amplitude. A discussion on
the rigorous meaning
of such an amplitude in terms of Thermal Field Theory,
as well as the explicit expression of the thermal corrections
can be found in \cite{GomezNicola:2002tn}. For our purposes here,
it is enough to say that the thermal amplitude 
can be projected into partial waves
that satisfy a generalized perturbative unitary relation:
 \begin{equation} 
\label{pertunitT}
\ima t_2 (s)= 0, \qquad \ima t_4 (s;\beta) = \sigma_T (S_0)
\left\vert t_2(s)\right\vert^2\ , \quad S_0>2m_\pi, \beta=1/T
\end{equation}
where
\begin{equation} \label{sigmaT}
\sigma_T (E)=\sigma(E^2)\left[1+\frac{2}{\exp({\beta\vert E
\vert/2})-1}\right]
\end{equation}
is the thermal two-particle
phase space in the c.o.m. frame. 
It is nothing but the phase space $\sigma(s)$  defined above
but corrected by the presence of 
two-pion states following a Bose-Einstein distribution.

In the dilute gas approximation it is natural to assume that 
this unitarity relation should be satisfied to all order, which 
thus leads to a straightforward thermal
generalization of the IAM \cite{Dobado:2002xf}. In the elastic approximation it is
again the IAM formula, but simply using the thermal amplitudes,i.e.,
replacing $t_4(s)$ by $t_4(s,\beta)$, calculated in \cite{GomezNicola:2002tn}.
Obviously, we recover the zero temperature unitarized 
amplitude when we make $T\rightarrow0$.
We recall again that the temperature dependence appears at this order through
 finite contributions to the loop functions, but not in the chiral parameters.
Thus, once we have parameters that generate the $\sigma$ and $\rho$ at $T=0$,
we can follow their thermal evolution by changing $T$.
Nevertheless, although the unitarization allows
to extend the ChPT applicability range,
the dilute approximation  implies that we can only
look at temperatures up to $T\simeq200\,\hbox{MeV}$.
Thus, in Fig2.b \cite{Dobado:2002xf} we can see the thermal evolution of
the $\sigma$ and  $\rho$ poles. Note that the width
of both resonances is growing with the temperature.
For the $\rho$ this is in good agreement with observations of the 
dilepton spectra produced in Heavy Ion Collisions.
For the $\sigma$ it could be interpreted as a signal
of chiral symmetry restoration.

\section{V. Unitarized ChPT and the large $N_c$ }

QCD is not perturbative below 1 or 2 GeV.
However, to understand many qualitative features of QCD and
also as a guiding line to organize calculations
we can use the large $N_c$ expansion \cite{largen}, 
even though the number of colors  is actually $N_c=3$.
The advantage of the large $N_c$ limit to study resonances is that 
$q\bar{q}$ states become bound states as $N_c\rightarrow\infty$. 
More quantitatively, their mass should remain almost constant
$M\simeq O(1)$, whereas
their decay width to two mesons should behave as $\Gamma\simeq O(1/N_c)$.
A similar behavior holds for glueballs decaying to two mesons.
In contrast, some multiquark states as $qq\bar{q}\bar{q}$ are expected to 
become unbound, i.e., to be part of 
 the meson-meson continuum \cite{Jaffe}.

The parameters in the ChPT Lagrangian
inherit the large $N_c$ dynamics of QCD.
In particular, 
the meson masses and decay constants behave as 
$f_\pi, f_K, f_\eta=O(\sqrt{N_c})$
and $M_\pi,M_K,M_\eta=O(1)$. In Table 6, we list
the large $N_c$ behavior of the chiral parameters \cite{chptlargen},
which is established in a model independent way.
Note however that there is an uncertainty
on the renormalization scale that corresponds to $N_c=3$, although
generically it should lie in the
$\mu\simeq 0.5-1\,$GeV range.

\begin{table}[hbpt]
\begin{tabular}{|c||c||c||c|}
\hline
  \tablehead{1}{r}{b}{Parameter\\
$\times 10^{-3}$} &
  \tablehead{1}{c}{b}{ChPT\\
\hspace*{.5cm}$\mu=770$ MeV\hspace*{.5cm}} &
  \tablehead{1}{c}{b}{IAM I\\
\hspace*{.5cm}$\mu=770$ MeV\hspace*{.5cm}}
&
  \tablehead{1}{c}{b}{Large $N_c$\\ \hspace*{.5cm}behavior\hspace*{.5cm}} 
\\
\hline
$2 L_1- L_2$
& $-0.6\pm0.2$
& $0.56\pm0.10$ 
& $O(1)$
\\
$L_2$
& $1.35\pm0.3$ 
& $1.21\pm0.10$ 
& $O(N_c)$\\
$L_3 $  &
 $-3.5\pm1.1$&
$-2.79\pm0.14$ 
& $O(N_c)$
\\
$L_4$
& $-0.3\pm0.5$& $-0.36\pm0.17$ 
& $O(1)$\\
$L_5$
& $1.4\pm0.5$& $1.4\pm0.5$ 
& $O(N_c)$
\\
$L_6$
& $-0.2\pm0.3$& $0.07\pm0.08$ 
&$O(1)$\\
$L_7 $  &
$-0.4\pm0.2$&
$-0.44\pm0.15$ &
$O(1)$
\\
$L_8$
& $0.9\pm0.3$& $0.78\pm0.18$ 
&$O(N_c)$\\
\hline
\end{tabular}
\caption{Different sets of chiral parameters ($\times10^{3}$).
For illustration, the 
ChPT and IAM I columns are repeated from Table 1. 
Other IAM sets give similar results. 
The last column shows the leading large $N_c$ behavior,
calculated from QCD. 
} 
\label{elesln}
\end{table}

Since we have built our unitarized amplitudes using
the completely renormalized ChPT expressions, we can 
now study the large $N_c$ behavior of the generated resonances
and get a hint on their nature. We will simply 
scale the ChPT parameters at $\mu=770\,$MeV  in the IAM amplitudes
fitted to the data, which therefore correspond to $N_c=3$.
As already commented, 
for {\it narrow} resonances the pole position
$\sqrt{s_{pole}}\simeq M-i\Gamma/2$
$M$ and $\Gamma$ being the mass and width of the resonance.
From the spectroscopic point of view,
a resonance could be a mixture of several states,
but comparing its large $N_c$ behavior
we could, in principle, elucidate the nature of the dominant
component.  Indeed, 
already in \cite{JA} it was shown, using unitarized meson-meson scattering
from a chiral resonance Lagrangian and only the s-loops unitarized with
an N/D method, that the lightest scalar resonances did not
behave as one would expect for $\bar{q}q$ states.
We will next show the results of our, still preliminary, study
with the unitarized ChPT renormalized amplitudes.
Of course, if most of the states of the
mixture disappears in the large $N_c$ limit,
a very small portion of a $q\bar{q}$ could become more patent
if we increase $N_c$ sufficiently. To avoid these effects we will  
present results for $N_c$ up to about 20.

\begin{figure}[hbpt]
\includegraphics[height=0.4\textheight]{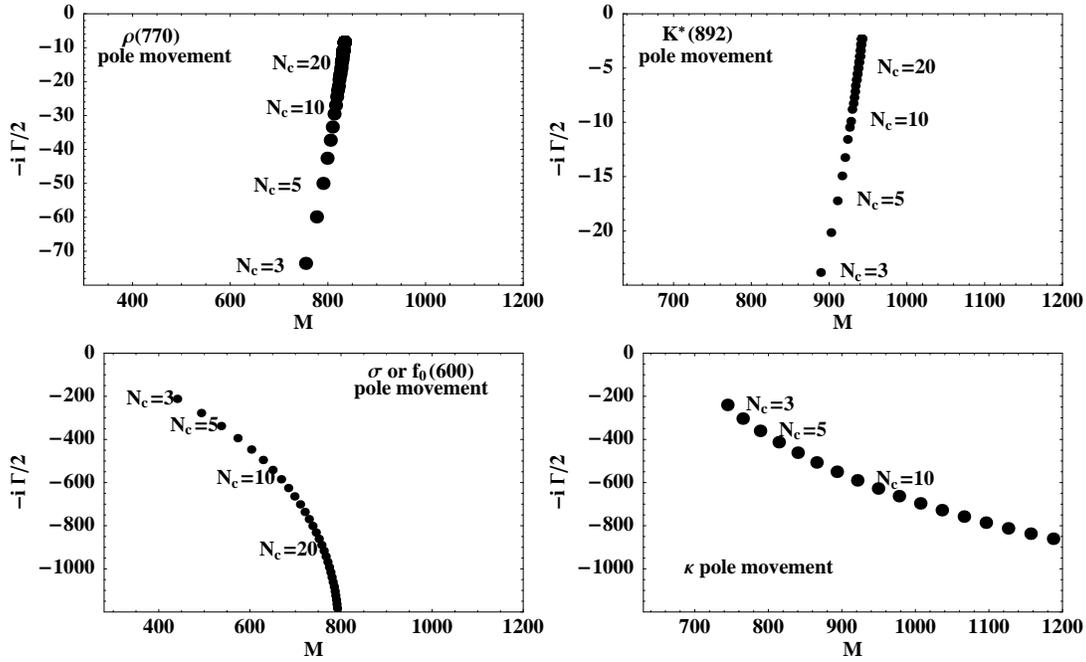}
\caption
{ Large $N_c$ dependence of the pole positions
in the lower half of the second Riemann $\sqrt{s}$ 
sheet of the meson meson scattering amplitude, obtained
from the unitarized one-loop Chiral Perturbation Theory calculation.
For each value of $N_c$ the pole is represented by a dot,
in different meson-meson
scattering channel. 
Note that the $\sigma$ and $\kappa$ behavior is opposite to that
of well know vector states as the $\rho$ and $K^*$.}  
\label{fig:largen}
\end{figure}

Thus, Fig.3 shows the large $N_c$ behavior of the 
poles in several channels of meson meson scattering.
Each dot corresponds to a different value of $N_c$.
First of all, we want to be sure that the method works
for well established $q\bar{q}$ states and so 
we turn to vector resonances.
Hence, on the top left, we represent the movement
of the $\rho(770)$ pole 
the $I,J=1,1$ channel of $\pi\pi$
scattering. On the top right we display the $K^*(982)$ movement
in the $I,J=1,1/2$ channel of  $\pi K$ scattering.
Remarkably, the IAM reproduces the 
expected $q\bar{q}$ behavior since
the mass of both vectors tends to a constant and
their width decreases as $1/N_c$. If we did not know it already,
this would be a strong hint that both states are indeed $q\bar{q}$
states. Let us remark that ChPT is built in terms of mesons,
not of quarks, but the QCD large $N_c$ dynamics  is correctly
inherited in  the $L_i$ values.

Let us then turn
to the controversial scalar resonances, subject of this workshop..
On the bottom left of Fig.3 we represent the movement of the pole 
commonly associated to the $\sigma$ and, on the right, that associated with
the $\kappa$.  Note that we keep the notation
$\sqrt{s_{pole}}\simeq M-i\Gamma/2$, although
 these poles are so far from the real axis (so wide)
that the interpretation in terms of mass and width
is no longer straightforward. However,
it is very clear that their behavior is completely at
variance with that expected for $q\bar{q}$ states. We can see that
in both cases, either $M$ or $\Gamma${\it  grow} as  $N_c$ is increased.
This behavior does not correspond
to a  $q\bar{q}$ or glueball nature. Other interpretations
should be invoked, although we want to
remark that the one that is becoming more widely accepted, namely,
 the four-quark state interpretation \cite{Jaffe}
(and also the two-meson molecules),
is able to accommodate the fact that these states become
the meson-meson continuum as $N_c$ grows.

\section{Conclusion and outlook}

In this talk I have reviewed the most recent developments
in the unitarization of the Chiral Perturbation Theory amplitudes,
paying particular attention to the Inverse Amplitude Method.
The outcome of this approach is of 
interest for meson spectroscopy since it can generate
 resonances, without including them in 
the ChPT Lagrangian, and
respecting simultaneously chiral symmetry and unitarity.

I have reviewed how unitarized meson-meson scattering
ChPT amplitudes provide a simultaneous description
of the low energy and resonant regions below 1.2 GeV,
generating the poles associated
to the $\rho,K^*, \sigma,\kappa,a_0$ and $f_0$ states.
This description respects the low energy chiral expansion up to 
next to leading order and the unitarized
fit leads to 
parameters compatible with those of standard ChPT.
Remarkably, it yields scattering lengths compatible with
higher order calculations and the most recent low energy data.
The amplitudes are completely renormalized
and  scale independent, thus avoiding any spurious dependence from
artificial scales.
We have also seen that some of the drawbacks, as crossing symmetry, seem
to be well under control, given the present experimental uncertainties.
The IAM is very simple to implement, and it can be systematically
extended to higher orders. The few unitarized
calculation to two loops indeed provide good descriptions 
of $\pi\pi$ scattering again with compatible parameters.
It has also been recently extended to channels with vanishing leading order
providing a satisfactory description of the $f_2(1250)$ resonance.
The description in terms of poles is robust, with the possible
exception of the $a_0$ (that could alternatively be interpreted as a cusp),
and these states
seem to be an unavoidable requirement of chiral symmetry
{\it and} unitarity. 

We have also shown how it can be generalized to thermal amplitudes,
which allowed us to study the temperature evolution of the $\sigma$
and $\rho$ resonances. 

Concerning the nature of these states, we have
presented a preliminary study of the large $N_c$
behavior of the poles. We have seen that 
the vectors generated with the IAM follow remarkably well the
expected $q\bar{q}$ behavior.
In contrast, the $\sigma$ and $\kappa$ states, behave
in a completely different way, ruling out a
$q\bar{q}$ or glueball interpretation. A $qq\bar{q}\bar{q}$
interpretation seems at least qualitatively adequate 
large $N_c$ evolution.

The large $N_c$ study is still preliminary 
and we are finishing the analysis of the $f_0$ and $a_0$ behavior,
which have larger uncertainties and a complicated 
interpretation due to 
the proximity of thresholds and possible mixings with
more massive multiplets 
\footnote{While preparing this work, an
study of the mass matrix of scalar resonances 
that survive in the large $N_c$ has appeared \cite{Cirigliano:2003yq}.
Claiming that the $f_0$ could be such a state and 
also the $a_0$ in one possible scenario}. We are also
estimating the errors due to the uncertainty in the 
renormalization scale where the large $N_c$ scaling is imposed.
Finally, we are presently looking directly
at the large $N_c$ to avoid using poles whose interpretation
for wide
resonances is delicate. Work is still in progress
in all these directions.

\begin{theacknowledgments}
First of all, I wish to thank the workshop organizers, 
specially A. Fariborz, for their  kind invitation 
and for their efforts to offer us such a nice atmosphere in Utica.  
I also wish to thank my collaborators A. Dobado, F. J. Llanes-Estrada
and specially A. G\'omez Nicola for his comments 
and his careful reading 
of the manuscript. I also 
thank A. Andrianov, D. Espriu, R. Jaffe,
 F. Kleefeld, M. Uehara and E. van Beveren
for their encouragement and interesting discussions .
Work supported by a Marie Curie fellowship, contract MCFI-2001-01155,
the Spanish CICYT projects, FPA2000-0956,
PB98-0782 and BFM2000-1326, the
CICYT-INFN collaboration grant 003P 640.15, and the 
E.U. EURIDICE network contract no. HPRN-CT-2002-00311.
\end{theacknowledgments}

\bibliographystyle{aipproc}   

\begin{thebibliography}{99}

\bibitem{chpt1}
S. Weinberg, Physica {\bf A96} (1979) 327.
J.~Gasser and H.~Leutwyler,
Annals Phys.\  {\bf 158} (1984) 142.
\bibitem{chpt2}
J.~Gasser and H.~Leutwyler,
Nucl.\ Phys.\ B {\bf 250} (1985) 465.


\bibitem{chptlargen}
A.~A.~Andrianov,
Phys.\ Lett.\ B {\bf 157}, 425 (1985).
A.~A.~Andrianov and L.~Bonora,
Nucl.\ Phys.\ B {\bf 233}, 232 (1984).
D.~Espriu, E.~de Rafael and J.~Taron,
Nucl.\ Phys.\ B {\bf 345} (1990) 22
[Erratum-ibid.\ B {\bf 355} (1991) 278].
S.~Peris and E.~de Rafael,
Phys.\ Lett.\ B {\bf 348} (1995) 539

\bibitem{Ecker:1988te}
G.~Ecker, J.~Gasser, A.~Pich and E.~de Rafael,
Nucl.\ Phys.\ B {\bf 321}, 311 (1989).


\bibitem{Truong:1988zp}
T.~N.~Truong,
Phys.\ Rev.\ Lett.\  {\bf 61} (1988) 2526.
\PRL{67}, (1991) 2260;
A. Dobado, M.J.Herrero and T.N. Truong, \PL{B235} (1990) 134.
A.~Dobado and J.~R.~Pelaez,
Phys.\ Rev.\ D {\bf 47} (1993) 4883.

\bibitem{Dobado:1996ps}
A.~Dobado and J.~R.~Pelaez,
Phys.\ Rev.\ D {\bf 56} (1997) 3057.

\bibitem{Nieves:1999bx}
J.~Nieves and E.~Ruiz Arriola,
Nucl.\ Phys.\ A {\bf 679} (2000) 57

\bibitem{Oller:1997ng}
J.~A.~Oller, E.~Oset and J.~R.~Pel\'aez,
Phys.\ Rev.\ Lett.\  {\bf 80} (1998) 3452;
Phys.\ Rev.\ D {\bf 59} (1999) 074001
[Erratum-ibid.\ D {\bf 60} (1999) 099906].
Phys.\ Rev.\ D {\bf 62} (2000) 114017.


\bibitem{Guerrero:1998ei}
F.~Guerrero and J.~A.~Oller,
Nucl.\ Phys.\ B {\bf 537} (1999) 459
[Erratum-ibid.\ B {\bf 602} (2001) 641].


\bibitem{GomezNicola:2001as}
A.~G\'omez Nicola and J.~R.~Pel\'aez,
Phys.\ Rev.\ D {\bf 65} (2002) 054009


\bibitem{Coimbra}
J.~R.~Pelaez and A.~Gomez Nicola,
AIP Conf.\ Proc.\  {\bf 660} (2003) 102
[arXiv:hep-ph/0301049].


\bibitem{Kpi}
V. Bernard, N. Kaiser, U.G. Mei{\it ss}ner, \PR{D43} (1991) 2757;
\NP{B357} (1991) 129;  \PR{D44} (1991) 3698.



\bibitem{BijnensKl4} G. Amor\'os, J. Bijnens and P. Talavera,
\NP{B602} (2001) 87.

\bibitem{BijnensGasser} J. Bijnens, G. Colangelo and J. Gasser,
\NP{B427} (1994) 427.


\bibitem{Amoros:2000mc}
G.~Amoros, J.~Bijnens and P.~Talavera,
Nucl.\ Phys.\ B {\bf 585} (2000) 293
[Erratum-ibid.\ B {\bf 598} (2001) 665].



\bibitem{pipidata}
S. D. Protopopescu {\em et al.}, Phys. Rev. {\bf D7}, (1973) 1279;
P. Estabrooks and A.D.Martin, Nucl.Phys.{\bf  B79}, (1974) 301. 
G. Grayer {\em et al.}, Nucl. Phys. {\bf B75}, (1974) 189. 
D. Cohen, Phys. Rev. {\bf D22}, (1980) 2595. 
W. Hoogland {\it et al.}, \NP{B126} (1977) 109.
M. J. Losty {\it et al.}, \NP{B69} (1974) 185.
 R. Mercer {\em et al.}, \NP{B32} (1971) 381.
 P. Estabrooks {\em et al.}, \NP{B133} (1978) 490.
 H. H. Bingham {\em et al.}, Nucl. Phys. {\bf B41} (1972) 1.
 S. L. Baker {\em et al.}, \NP{B99} (1975) 211.
D. Aston {\em et al.}
Nucl. Phys. {\bf B296} (1988) 493.
 D. Linglin {\em et al.}, \NP{B57} (1973) 64 .
S.~Pislak {\it et al.}  [BNL-E865 Collaboration],
Phys.\ Rev.\ Lett.\  {\bf 87} (2001) 221801.
P. Estabrooks and A.D. Martin, \NP{B79} (1974) 301
C.D. Froggat and J.L. Petersen,\NP{B129}(1977)89.



\bibitem{Hannah:1997ux}
T.~Hannah,
Phys.\ Rev.\ D {\bf 55} (1997) 5613


\bibitem{Nieves:2001de}
J.~Nieves, M.~Pavon Valderrama and E.~Ruiz Arriola,
Phys.\ Rev.\ D {\bf 65} (2002) 036002


\bibitem{bijnens} J. Bijnens {et al.}, \PL{B374} (1996) 210; 
\NP{B508} (1997) 263.

\bibitem{Dobado:2001rv}
A.~Dobado and J.~R.~Pelaez,
Phys.\ Rev.\ D {\bf 65} (2002) 077502
[arXiv:hep-ph/0111140].

\bibitem{newsigma}
R.L. Jaffe, \PR{D15} 267 (1977); \PR{D15}, 281 (1977).
E. van Beveren {\it et al.} \ZP{C30}, 615 (1986).
R.~Kaminski, L.~Lesniak and J.~P.~Maillet,
Phys.\ Rev.\ D {\bf 50} (1994) 3145.
M.~Harada, F.~Sannino and J.~Schechter,
Phys.\ Rev.\ D {\bf 54} (1996) 1991
R.~Delbourgo and M.~D.~Scadron,
Mod.\ Phys.\ Lett.\ A {\bf 10} (1995) 251.
S.~Ishida {\it et al.}, 
Prog.\ Theor.\ Phys.\  {\bf 95} (1996) 745.
N.~A.~Tornqvist and M.~Roos,
Phys.\ Rev.\ Lett.\  {\bf 76} (1996) 1575.
S. Ishida {\it et al}, Prog. Theor. Phys. 98,621 (1997).
D. Black, A. H. Fariborz, F. Sannino, J. Schechter. 
\PR{D58}:054012,1998. 
E.~van Beveren and G.~Rupp,
Eur.\ Phys.\ J.\ C {\bf 22} (2001) 493
E.~van Beveren and G.~Rupp,
arXiv:hep-ph/0201006.

\bibitem{JA}
J.~A.~Oller and E.~Oset,
Phys.\ Rev.\ D {\bf 60} (1999) 074023

\bibitem{charm}E791 Collaboration,\PRL{86},(2001) 770.
E.~M.~Aitala {\it et al.}  [E791 Collaboration],
Phys.\ Rev.\ Lett.\  {\bf 89} (2002) 121801
C. Gobel for the E791 Collab. hep-ex/0012009.



\bibitem{Oller:1997ti}
J.~A.~Oller and E.~Oset,
Nucl.\ Phys.\ A {\bf 620} (1997) 438
[Erratum-ibid.\ A {\bf 652} (1999) 407]


\bibitem{PDG}  K. Hagiwara {\it et al.}, Phys. Rev. {\bf D 66}, 010001 (2002).


\bibitem{Uehara:2002nv}
M.~Uehara,
arXiv:hep-ph/0204020.

\bibitem{GomezNicola:2002tn}
A.~Gomez Nicola, F.~J.~Llanes-Estrada and J.~R.~Pelaez,
Phys.\ Lett.\ B {\bf 550} (2002) 55

\bibitem{Dobado:2002xf}
A.~Dobado, A.~Gomez Nicola, F.~J.~Llanes-Estrada and J.~R.~Pelaez,
Phys.\ Rev.\ C {\bf 66} (2002) 055201


\bibitem{largen}
G.~'t Hooft,
Nucl.\ Phys.\ B {\bf 72} (1974) 461.
C.~Rosenzweig, J.~Schechter and C.~G.~Trahern,
Phys.\ Rev.\ D {\bf 21} (1980) 3388.
E.~Witten,
Annals Phys.\  {\bf 128} (1980) 363.

\bibitem{Jaffe} R. L. Jaffe, Proceedings of the Intl. Symposium
on Lepton and Photon Interactions at High Energies. Physikalisches Institut, University of Bonn (1981) . ISBN: 3-9800625-0-3 

\bibitem{Cirigliano:2003yq}
V.~Cirigliano, G.~Ecker, H.~Neufeld and A.~Pich,
arXiv:hep-ph/0305311.


\end{thebibliography}

\end{document}